\documentclass[conference]{IEEEtran}

\usepackage{amssymb}
\usepackage{amsmath}
\usepackage{graphicx}
\usepackage{color}	
\usepackage{cite}

\newcommand{\GF}{\mathrm{GF}}

\begin{document}

\title{Ultra Low Complexity Soft Output Detector for Non-Binary LDPC Coded Large MIMO Systems}

\author{\IEEEauthorblockN{Puripong Suthisopapan$^{\dagger}$, Anupap Meesomboon$^{\dagger}$, Kenta Kasai$^{\ddagger}$,  
and Virasit Imtawil$^{\dagger}$}
\IEEEauthorblockA{$^{\dagger}$Dept. of Electrical Engineering, Faculty of Engineering, 
Khon Kaen University, Thailand 40002\\
$^{\ddagger}$Dept. of Communications and Integrated Systems, Tokyo Institute of Technology, 152-8550 Tokyo, Japan  \\
Email: mr.puripong@gmail.com, anupap@kku.ac.th, kenta@comm.ss.titech.ac.jp, virasit@kku.ac.th}
}

\maketitle


\begin{abstract}
The theoretic results of MIMO capacity tell us that 
the higher the number of antennas are employed, the higher the transmission rate is.
This makes MIMO systems with hundreds of antennas very attractive 
but one of the major problems that obstructs such large dimensional MIMO systems from the practical realization
is a high complexity of the MIMO detector.
We present in this paper the new soft output MIMO detector based on matched filtering that can be applied
to the large MIMO systems which are coded by the powerful non-binary LDPC codes.
The per-bit complexity of the proposed detector is just 0.28$\%$ to that of low complexity soft output MMSE detector
and scales only linearly with a number of antennas.
Furthermore, the coded performances with small information length 800 bits are within 4.2 dB from the associated MIMO capacity.
\end{abstract}

%
\IEEEpeerreviewmaketitle

\section{Introduction}
Over recent years, the multiple input multiple output (MIMO) systems 
which employ tens to hundreds transmitted/received antennas
have been a subject of considerable interest \cite{large_MIMO_krylov,large_MIMO_SSK,large_MIMO_Tabu}.
Without additional bandwidth and transmitted power requirements,
such large MIMO systems can provide higher transmission rate simultaneously with better reliability.
The optimal maximum likelihood (ML) detector is known to be infeasible for large MIMO systems
since its complexity grows exponentially with the number of transmitted antennas and the modulation order.
Therefore, the research on designing the low complexity detector for large MIMO systems is garnering much attention \cite{large_mimo_detector}.

In this paper, we intend to deal with the coded large MIMO systems in which the operating region, i.e., BER of $10^{-5}$, is in low SNRs.
We thus  consider only the classical linear detectors such as matched filtering (MF) or  minimum mean square error (MMSE) techniques
since the detection performance of these linear detection schemes at low SNRs seems to be near optimal \cite{massive_mimo}.
Most importantly, these kinds of detectors have very low complexity and can be applied to large MIMO system.

Although practical coded large MIMO systems have been proposed in the literature, 
the performance gap to the capacity remains large \cite{coded_large_mimo1,coded_large_mimo3}.
For example, the performance of turbo coded large MIMO systems with likelihood ascent search (LAS) detection \cite{coded_large_mimo1}
is more than $7.5$ away from its associated capacity.
We have recently shown in our previous study that 
non-binary LDPC coded large MIMO systems with soft output MMSE detection and small code length
can be employed to reduce the remaining gap to just 3.5 dB with affordable complexity \cite{my_arxiv}.
However, MMSE detector needs to calculate the inverse of large and dense matrix 
and this calculation is known to has the cubic complexity with a number of received antennas \cite{complexity_cal}.
To avoid such matrix inversion, 
we propose in this paper the soft output MIMO detector based on MF technique for large MIMO systems which are coded by non-binary LDPC codes.
The complexity analysis provided in this paper indicates that
the detection complexity of the proposed detector is very low, i.e., 
it is just 0.28$\%$ to that of the soft output MMSE detection
while the loss in coded performance, compared to the case of employing MMSE detection, is marginal especially at very low SNRs.

The rest of this paper is organized as follows. 
We first describe the system model in Section II. 
In Section III, the proposed soft output MF-based detector is presented. 
In Section IV, we present the performance of the non-binary LDPC coded large MIMO systems which employ the proposed detector.
In Section V, the complexity analysis of the proposed detector is provided.
Finally, the conclusions are given.  

\section{System Model}
\begin{figure}[htb]
\centering
\includegraphics[scale=0.65]{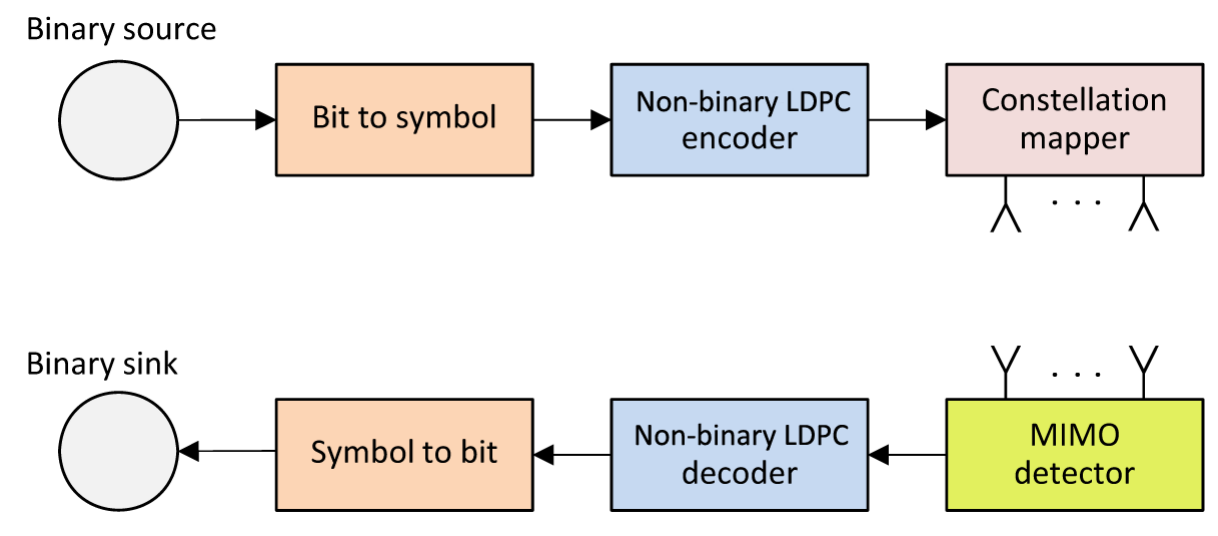}
\caption{ The MIMO system concatenated with non-binary LDPC codes.}
\label{NBLDPC_coded_MIMO}
\end{figure}

In this study, $N_t \times N_r$ MIMO system represents the MIMO system with $N_t$ transmit antennas and $N_r$ receive antennas.
Let $\mathbb{A}^{M}$ be the complex modulation constellation of size $M=2^p$ where $p$ represents bit(s) per modulated symbol.
Figure \ref{NBLDPC_coded_MIMO} shows the MIMO system concatenated with non-binary LDPC codes of rate $R=K/N$.
In this paper, non-binary LDPC codes defined over $\GF(2^8)$ are chosen because 
the excellent performance can be achieved with simple code structure \cite{my_arxiv,nb_binary_image}, 
i.e., ultra sparse regular parity-check matrix.

At the transmitter side, a bit to symbol mapper maps 
a group of 8 information bits to a non-binary symbol in $\GF(2^8)$.
The non-binary LDPC encoder then encodes the stream of $K$ symbols into a codeword of length $N$ symbols.
Each coded symbol in $\GF(2^8)$ is then mapped to a group of $q=$ 8$/p$ modulated symbols by a constellation mapper.
At each time instant (each channel use), 
the transmitter simultaneously sends $N_t = K_tq$ modulated symbols in parallel 
through $N_t$ transmit antennas where $K_t$ is a number of coded symbols per each transmission.
Let $\mathbf{s} = [s_1,s_2,\ldots,s_{N_t}]^{T}\in\mathbb{C}^{N_{t}}$ be the transmit signal vector.
Each entry $s_i , \forall i \in \lbrace1,\ldots,N_t\rbrace $ taken from $\mathbb{A}^M$ must satisfy the component-wise energy constraint $\mathrm{E}[\|s_i\|^2]= E_s/N_{t}$ 
where $E_s$ is the total transmitted power, 
$\parallel\cdot\parallel^2$ denotes the squared Euclidean norm, 
and  $\mathrm{E}[\cdot]$ denotes the expectation.


Considering a baseband discrete time model for uncorrelated flat fading MIMO channel,
the received vector $\mathbf{y} = [y_1,y_2,\ldots,y_{N_r}]^{T}\in\mathbb{C}^{N_{r}}$ 
of the $N_t \times N_r$ MIMO system is given by \cite{coded_large_mimo1}
\begin{equation}
\label{model_eq}
\mathbf{y} = \mathbf{H}\mathbf{s} + \mathbf{n}.
\end{equation}
The matrix 
$\mathbf{H} = [\mathbf{H}_1 \mathbf{H}_2 \ldots \mathbf{H}_{N_t}] \in\mathbb{C}^{N_{r} \times N_{t}} $ 
denotes the channel fading matrix 
whose entry $h_{kj}$ is assumed to be complex Gaussian random variable 
with zero mean and unit variance.
The vector $\mathbf{n} = [n_1,n_2,\ldots,n_{N_r}]^{T}\in\mathbb{C}^{N_{r}}$ is a noise vector 
whose entry is a complex white Gaussian noise with zero mean and variance $\sigma^{2}_{n}$ per real component.
The MIMO detector performs detection and produces the prior probabilities (soft output) for non-binary LDPC decoder.
After the decoder has all information for $N$ coded symbols, 
the decoder performs decoding and then provides the estimated non-binary symbols (hard output).
These estimated symbols are finally demapped to a sequence of estimated information bits.
In this paper, the channel matrix $\mathbf{H}$ is assumed to be known at the receiver.

Since each entry of $\mathbf{H}$ has unit variance, 
the average signal energy per receive antenna is $E_s$.
We follow the convention that $N_0/2 = \sigma^{2}_{n}$ to define the signal to noise ratio.
In this setting, the average signal to noise ratio (SNR) per receive antenna, 
denoted by $\gamma$, is given by \cite{cap_approach1}
\begin{equation}
\label{SNR_eq}
\gamma = \frac{E_s}{N_0} = \frac{E_s}{2\sigma^{2}_{n}}.
\end{equation}

The spectral efficiency (net information rate per unit bandwidth) of coded MIMO system with spatial multiplexing technique 
is $pRN_t$ where $p$ represents a number of bits per modulated symbol and $R$ is code rate \cite{cap_approach1}.
With a perfect $\mathbf{H}$ at the receiver side, the ergodic MIMO capacity is given by \cite{mimo2}
\begin{equation}
\label{C_eq}
C = \mathrm{E}\left[ \log_2\det\left(\mathbf{I}_{N_r}+ \left(\gamma/N_t\right)\mathbf{H}\mathbf{H}^\mathsf{H}\right) \right],
\end{equation}
where the superscript $\mathsf{H}$ denotes the Hermitian transpose operator, $\det$ denotes the determinant 
and $\mathbf{I}_{N_r}$ is the identity matrix of size $N_r \times N_r$.
The unit of spectral efficiency and capacity is bits/sec/Hz (bps/Hz).

\section{MIMO Detector based on Matched Filtering}
Before going to describe the proposed detector, 
let us introduce a class of low complexity linear detection known as MMSE detector.
For the MMSE detector, it is needed to compute the weight matrix $\mathbf{W}$ which is given below \cite{BP_MIMO},
\begin{equation}
\label{mmse}
\mathbf{W}_{\textrm{mmse}} = \left( \mathbf{H}\mathbf{H}^\mathsf{H}  + \frac{N_0}{E_s/N_t} \mathbf{I}_{N_r} \right)^{-1} \mathbf{H}^{\mathsf{H}}.
\end{equation}
The MMSE detector then estimates $\hat{s}_i$ of the transmitted symbol on $i$th antenna
by multiplying the received vector $\mathbf{y}$ with the $i$th row of $\mathbf{W}_{\textrm{mmse}}$,
\begin{equation}
\label{estimation}
\begin{array}{ll}
  \hat{s}_i  &= \mathbf{W}_{\textrm{mmse},i}\mathbf{y}\\
\end{array} 
\end{equation}
where $\mathbf{W}_{\textrm{mmse},i}$ is the $i$th row of  $\mathbf{W}_{\textrm{mmse}}$.
Finally, $\hat{s}_i$ is sliced to obtain the hard estimation.

For large MIMO systems,
a major problem of MMSE detection is the matrix inversion which has cubic time complexity \cite{complexity_cal}.
In this paper, we avoid the matrix inversion by considering MF-based detection.
Although the MF technique for MIMO detection is well known in literature
but we present in this paper a novel soft output generation from MF-based detection for non-binary LDPC decoder 
which has not yet reported elsewhere.

%

\subsection{Detection based on Matched Filtering}
A baseband discrete time channel model given in (\ref{model_eq})
can be equivalently rewritten as 
\begin{equation}
\label{model2_eq}
\mathbf{y} = \mathbf{H}_1 s_1 + \mathbf{H}_2 s_2 + \cdots + \mathbf{H}_{N_t} s_{N_t} + \mathbf{n}.
\end{equation}
To estimate $\hat{s}_k$ of the transmitted symbol on the $k$th antenna by
the concept of matched filtering,
the interference streams in (\ref{model2_eq}) are treated as noise,
\begin{equation}
\label{assump_eq}
\mathbf{y} = \mathbf{H}_k s_k + \underbrace{ \sum\limits_{i=1,i \neq k}^{N_t}\mathbf{H}_i s_i + \mathbf{n} }_{\textrm{noise}}.
\end{equation}
Then, the MIMO detector based on MF technique
estimates $\hat{s}_k$ of the transmitted symbol on the $k$th antenna
by multiplying the received vector $\mathbf{y}$ with weight matrix 
$\mathbf{W}_k =\frac{\mathbf{H}_k^\mathsf{H}}{\mathbf{H}_k^\mathsf{H}\mathbf{H}_k}$,
\begin{equation}
\label{detect_eq}
\begin{array}{ll}
   \hat{s}_k&= \mathbf{W}_k\mathbf{y},\\
&= s_k + \mathbf{W}_k\sum\limits_{i=1,i \neq k}^{N_t}\mathbf{H}_i s_i + \mathbf{W}_k\mathbf{n}.
\end{array}
\end{equation}
In contrast of the conventional MF detection, 
we make a little modification by introducing the term $\frac{1}{\mathbf{H}_k^\mathsf{H}\mathbf{H}_k}$ into weight matrix.
After slicing $\hat{s}_k$,
we can obtain the estimated symbol transmitted from the $k$th antenna.
It is known that the detection performance (uncoded performance) of match filtering is very poor 
since there is no cancellation of the interference terms (the second term in R.H.S of (\ref{detect_eq})).
However, matched filtering is near-optimum if $\mathbf{n}$ is dominant \cite{massive_mimo}, i.e. at low SNRs. 


\subsection{Generation of Soft Output}

Assuming all streams and noise are statistical independence and Gaussian distributed,
the proposed method for generating soft output from MF-based detection will be described as follows.
Following the definition of the post detection signal to interference plus noise ratio ($\mathrm{SINR}$) given in \cite[p.~358]{soft_out}, 
the $\mathrm{SINR}$ of the proposed detection for $k$th stream, denoted by $\delta_k$, can be expressed as
\begin{equation}
\label{SINR_eq}
\begin{array}{ll}
   \delta_k& = \frac{\textbf{$\mathrm{E}[\Vert s_k \Vert^2]$}}
   {\textbf{$\mathrm{E}[\sum\limits_{i=1,i \neq k}^{N_t} \Vert \mathbf{W}_k\mathbf{H}_i s_i \Vert^2]$}
   ~+~\textbf{$\mathrm{E}[\Vert \mathbf{W}_k\mathbf{n} \Vert^2]$}},\\
 & = \frac{E_s/N_t}{E_s/N_t\sum\limits_{i=1,i \neq k}^{N_t}\|\mathbf{W}_k\mathbf{H}_i\|^{2}~+~\left(2\sigma^{2}_{n}\|\mathbf{W}_k\|^{2}\right)},\\
 & = \frac{E_s/N_t}{\Delta_k}. \\
\end{array}
\end{equation}
The denominator, denoted by $\Delta_k$, given in (\ref{SINR_eq}) can be approximated by a Gaussian random variable.
Based on this approximation, 
we set $\sigma^{2}_{k} = \Delta_k$ 
and the soft output which exactly is the likelihood of $\hat{s}_k$ conditioned on $s \in \mathbb{A}^{M}$ is as follows
\begin{equation}
\label{Prob_eq}
\mathrm{P}\left( \hat{s}_k \mid s \right) =  \frac{1}{\sqrt{2\pi\sigma^{2}_{k}}}~ \text{exp} \left( -\frac{1}{2\sigma^{2}_{k}}\|\hat{s}_k - s\|^{2}\right).
\end{equation}

\begin{figure}[htb]
\centering
\includegraphics[scale=0.64]{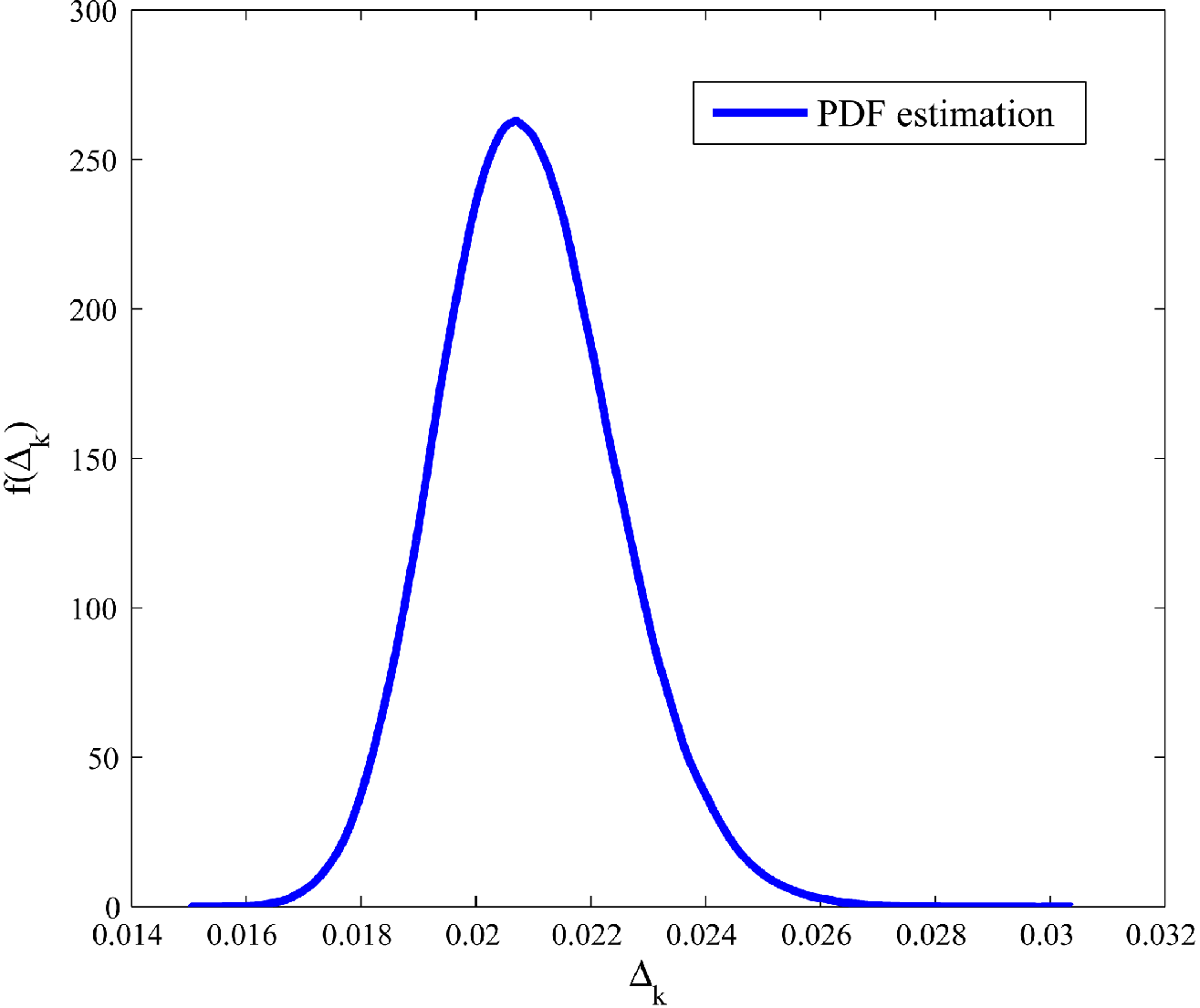}
\caption{The $\mathrm{pdf}$ estimation of $\Delta_k$ from $200 \times 200$ MIMO channel with BPSK modulation at SNR per receive antenna $\gamma$ = -2 dB.}
\label{PDF_estimation}
\end{figure}

To justify that $\Delta_k$ distributes like Gaussian,
we estimate the probability density function ($\mathrm{pdf}$) of random variable $\Delta_k$ 
by mean of Kernel density estimation \cite{prob_book}.
By using 10,000 channel realizations, 
it is obviously seen from Fig. \ref{PDF_estimation} 
that the $\mathrm{pdf}$ of $\Delta_k$ closely agrees with the shape of Gaussian distribution.
The mean just deviates from zero by 0.0209.
Therefore, we conclude that the Gaussian approximation of $\Delta_k$ described above is reasonable.

\subsection{Simplification}
To reduce the computation complexity of detection,
we observe that $\mathrm{E}\left[\mathbf{H}_k^\mathsf{H}\mathbf{H}_k\right] \approx N_r$.
Thus, the computation of $\mathbf{W}_k$ can be simplified as follows
\begin{equation}
\label{reduc1}
\begin{array}{ll}
\mathbf{W}_k & = \mathbf{H}_k^\mathsf{H} \left( 1/N_r \right) .\\
\end{array}
\end{equation}
With this simplification,
the matrix multiplication $\mathbf{H}_k^\mathsf{H}\mathbf{H}_k$ whose an complexity is $\mathcal{O}(N_r)$ \cite{matrix_computation} can be reduced to just a constant term.

To reduce the computational complexity of soft output generation,
we first observe that $\mathrm{E}\left[ \|\mathbf{W}_k\|^{2} \right] \approx 1/N_r$. 
Then, we assume that the power of noise term is dominant, i.e., $2\sigma^{2}_{n} \gg E_s$.
With this assumption, we have the following expression
\begin{align*}
E_s/N_t\sum\limits_{i=1,i \neq k}^{N_t}\|\mathbf{W}_k\mathbf{H}_i\|^{2} \ll  2\sigma^{2}_{n}\|\mathbf{W}_k\|^{2}.
\end{align*}
The above expression is valid when $\gamma$ is low, e.g., $\gamma$ = -5 dB,
since $1/N_t\sum\limits_{i=1,i \neq k}^{N_t}\|\mathbf{W}_k\mathbf{H}_i\|^{2} < \|\mathbf{W}_k\|^{2}$.
Therefore, the computation of variance $\sigma^2_k$ for soft output generation
defined in (\ref{Prob_eq}) can be reduced to a constant  term
\begin{equation}
\label{reduc2}
\begin{array}{ll}
\sigma^2_k & = E_s/N_t\sum\limits_{i=1,i \neq k}^{N_t}\|\mathbf{W}_k\mathbf{H}_i\|^{2}~+~\left(2\sigma^{2}_{n}\|\mathbf{W}_k\|^{2}\right),\\
 & \approx  2\sigma^{2}_{n}\|\mathbf{W}_k\|^{2},\\
 & \approx 2\sigma^2_n/N_r.
\end{array}
\end{equation}
The variance $\sigma^2_k$ for soft output generation is now independent of $k$ and can be easily pre-computed.

\section{Numerical Results}
In this section, the numerical results of the uncoded and the coded large MIMO systems are presented.
The non-binary LDPC codes defined over $\GF(2^8)$ with regular parity-check matrix of column weight two are deployed as channel code.
The decoder is implemented by FFT-based belief propagation algorithm \cite{FFT_BP} with maximum iteration 200.
The BPSK modulation is employed for transmission.
We shall refer to non-binary LDPC coded large MIMO system with the proposed MF-based detector 
as the proposed system for simplicity.

\begin{figure}[htb]
\centering
\includegraphics[scale=0.60]{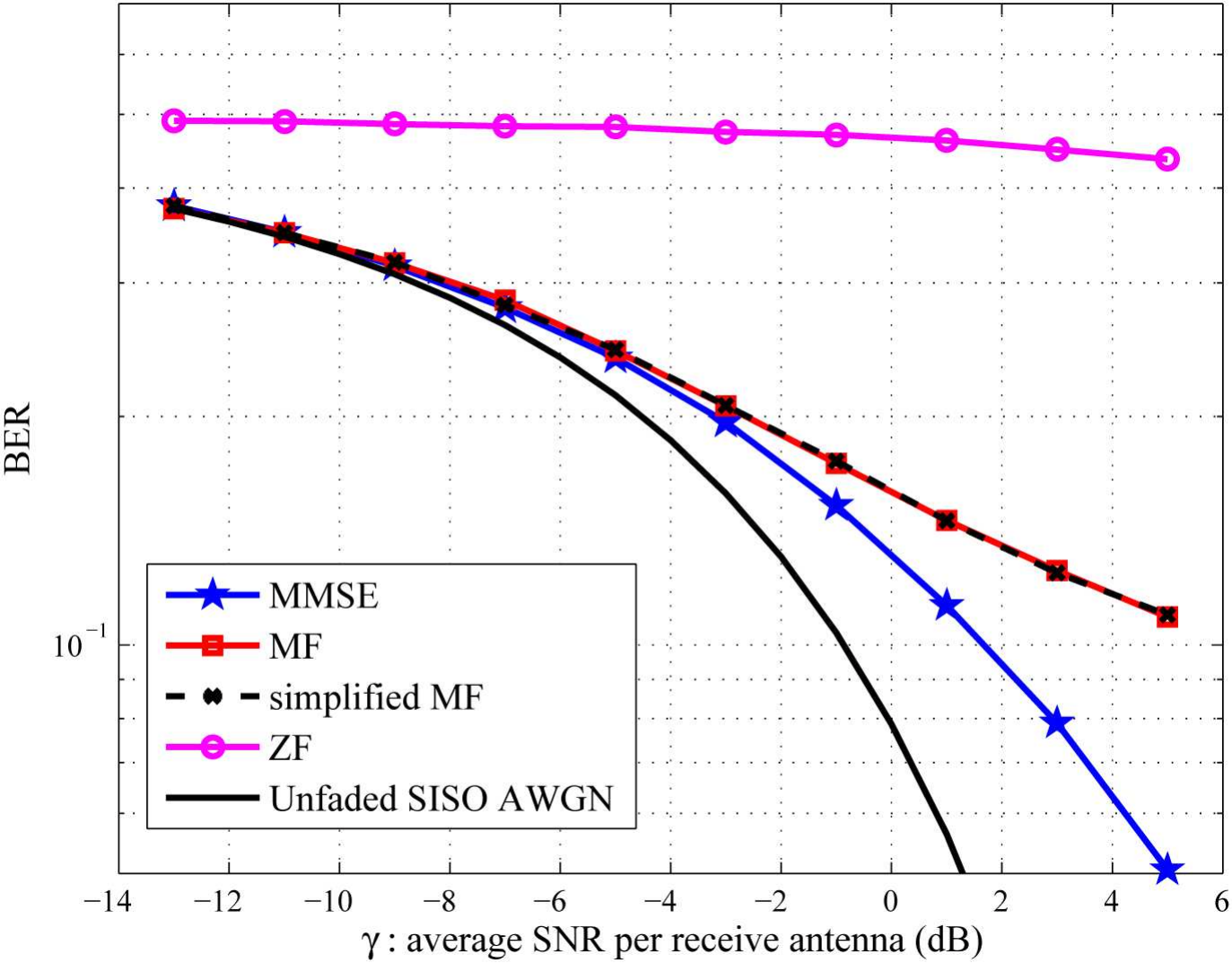}
\caption{Uncoded performance of MMSE, ZF, MF, and simplified MF detectors in $200\times200$ MIMO system with BPSK modulation.}
\label{uncoded_BER}
\end{figure}

Figure \ref{uncoded_BER} displays the uncoded performance of various low complexity linear detection schemes at low SNRs.
The uncoded peformance of single input single output (SISO) unfaded AWGN-BPSK system is also shown in the figure
to represent the approximated lower bound for ML performance \cite{large_MIMO_Tabu}.
Major observations can be summarized as follows :

$\bullet$ There is no performance difference between MF detection and its corresponding simplified version.

$\bullet$ The performance of MF detection is seriously degraded comparing to that of MMSE detection when $\gamma$ gets higher.

$\bullet$ The performance of MMSE and MF detections at low SNRs achieves near the performance of SISO unfaded AWGN.
Thus, both are near optimal in low SNR region.

$\bullet$ The zero forcing (ZF) detection performs very poor in low SNRs
so we do not pay the attention to this type of detector.

It is thus intuitive to think that 
the proposed MF-based detector would be useful when operating in low SNR region, i.e. at near MIMO capacity.

\begin{figure}[htb]
\centering
\includegraphics[scale=0.54]{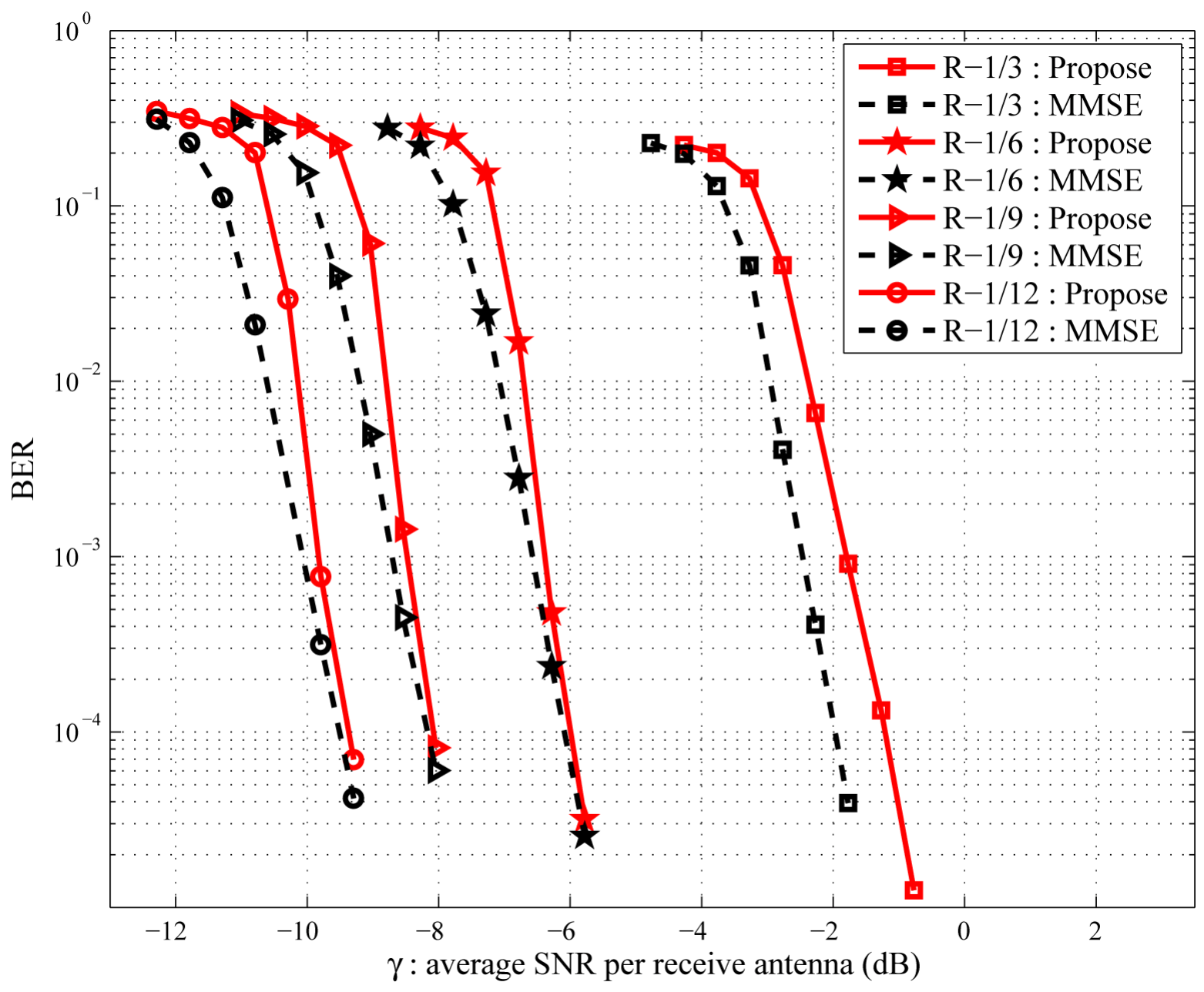}
\caption{BER of the non-binary LDPC coded $200 \times 200$ MIMO systems with BPSK modulation. 
``Propose" means the non-binary LDPC coded system with the proposed soft output MF-based detection and ``MMSE" represents the non-binary LDPC coded system with soft-output MMSE detection.}
\label{BER_result}
\end{figure}

Figure \ref{BER_result} shows the performance comparison between the proposed system 
and the non-binary LDPC coded system with soft output MMSE detector when the number of transmitted antennas is $N_t=200$.
The generation of soft outputs from MMSE detector are given in \cite{Eq_AWGN}.
The code rates of non-binary LDPC codes are ranging from 1/12 to 1/3 and the information length is 800 bits ($K=100$ symbols).
We select the low rate channel codes 
since we expect that the operating region, e.g. BER of $10^{-4}$, will be occurred in low SNRs 
at which the proposed detection is near optimal.
For $R=1/3$, the performance of coded system with MMSE detector is better than 
that of the proposed system by about 0.8 dB.
This is because the uncoded performance of MF-based detector is worse than that of MMSE detector by roughly 2 dB
as shown in the previous figure.
The performance difference between two coded systems (one with MMSE and another one with MF) is vanished 
for $R=1/6,1/9$ and $1/12$
since, in low SNRs, both detectors provides almost the same detecting performance.
Note that the codes with $R < 1/3$ are constructed by the instruction given in \cite{MRNB}.

\begin{figure}[htb]
\centering
\includegraphics[scale=0.62]{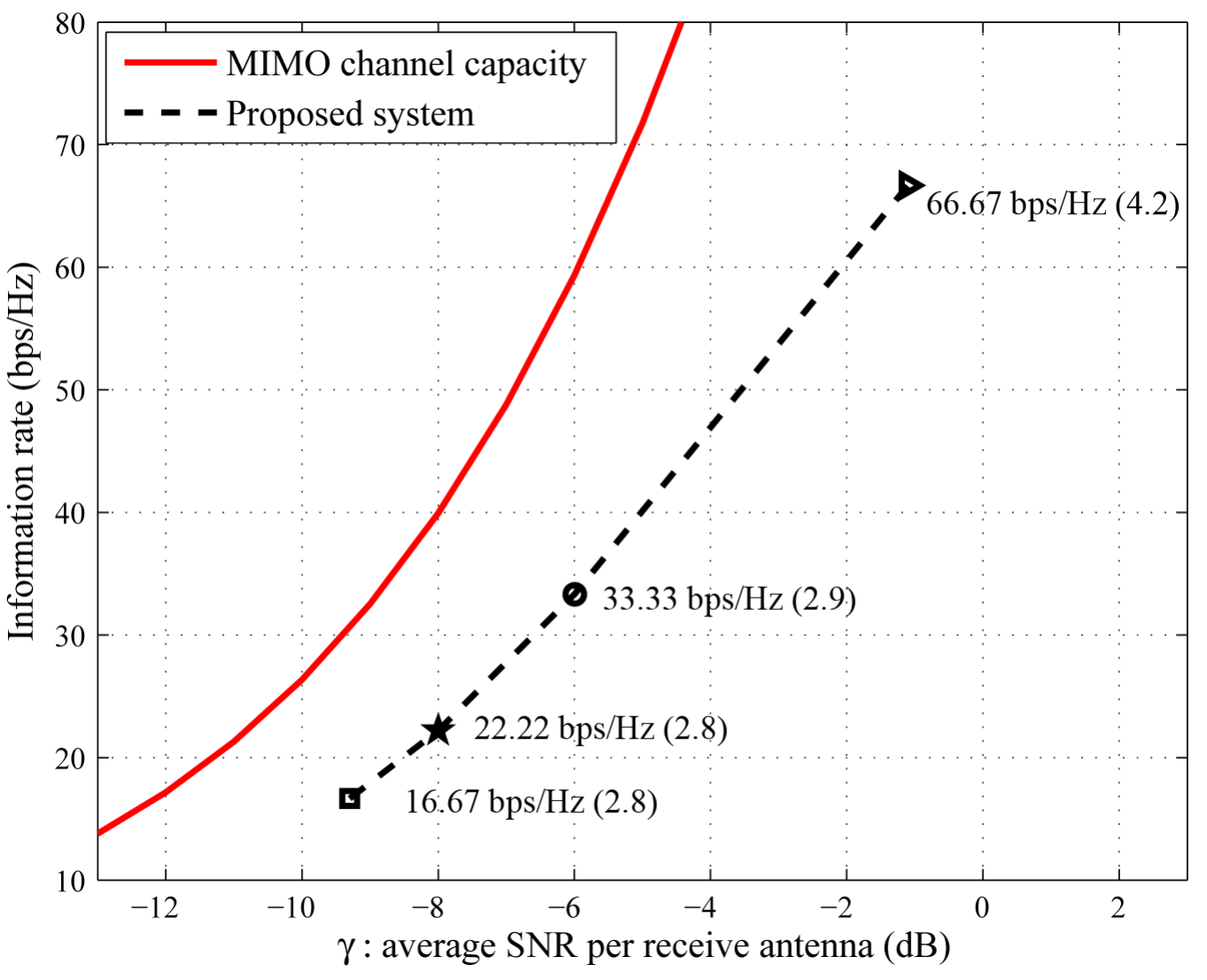}
\caption{Spectral efficiency of the proposed coded $200 \times 200$ MIMO systems.
The number in parentheses represents a gap to the capacity.
}
\label{spectral_results}
\end{figure}

Figure \ref{spectral_results} illustrates the  performance of the proposed system in terms of spectral efficiency.
We plot the SNR points of the proposed system required to reach the BER of $10^{-4}$.
For spectral efficiency below 66.67 bps/Hz, the proposed system can operate within 4.2 dB from the associated MIMO capacity.
Thanks to the large $N_t$, even we code the system with low rate non-binary LDPC codes but the high data rate still can be obtained.
For example, 1 Gbps over bandwidth of 50 MHz can be obtained from $R=1/9$ coded $200 \times 200$ MIMO-BPSK systems.
More interestingly, 
the reliable communications with very high spectral efficiency, about 16 bps/Hz, and extremely low SNR, roughly -10 dB, 
are unprecedented in literature.

\section{Complexity Analysis}
We will show in this section through complexity analysis that the performance loss from employing the proposed detector
is marginal when one looks at the complexity reduction.
Following \cite[p.~18]{matrix_computation,complexity_cal}, 
the complexity of complex matrix calculation is expressed in terms of complex floating point operations (flops).
The number of flops for some mathematical operations is given in Table \ref{matrix_comp}.
We note that the detection for coded MIMO systems comprises of two steps : 1) detection and 2) soft output generation.
\begin{table}[htb]
\setbox0\hbox{\verb/\documentclass/}%
\caption{The number of flops for some operations \cite{complexity_cal} where 
 $a,b \in \mathbb{R}$, $\mathbf{a},\mathbf{b} \in \mathbb{C}$, and $\mathbf{c},\mathbf{d} \in \mathbb{C}^{N_r \times 1}$}
\label{matrix_comp}
\begin{center}
\begin{tabular}{| c | c |  c | }
 \hline
 Operation (name) & Operation (math.) &  Number of flops \\
  \hline
Real multiplication & $ab$ & 1 \\
Complex multiplication & $ \mathbf{a}\mathbf{b}$ & 3 \\
Real addition & $a+b$ & 1 \\
Complex addition & $\mathbf{a} +\mathbf{b}$ & 1 \\
Inner product & $\mathbf{c^{\mathsf{H}}}\mathbf{d}$ & $4N_r-1$ \\
Scalar-vector multiplication & $\mathbf{a} \cdot a$ & $N_r$ \\
  \hline
\end{tabular}
\end{center}
\end{table}

\subsection{Proposed Soft Output Detector}
The estimation of $\hat{s}_k$ as defined in (\ref{detect_eq}) needs two steps : 
1) Computing the weight matrix  $\mathbf{W}_k$ according to (\ref{reduc1}) which requires $N_r$ flops.
2) Multiplying $\mathbf{W}_k$ with received vector $\mathbf{y}$ which requires $4N_r-1$ flops.
Since we have $N_r$ received antennas, 
the total flops used for detection are $5N_r^2 - N_r$.
Thus, the per-bit complexity is $\mathcal{O}(N_r)$.
To obtain the soft output defined in (\ref{Prob_eq}),
we need one flop for computing squared euclidean norm,  one  flop for subtraction, $3$ flops for multiplication of real constant, 
and assuming $50$ flops for exponential calculation \cite[p.193]{numerical_comp}.
Each transmitted symbol has $M$ levels to be calculated and we have $N_t = N_r$ transmitted symbols.
Therefore, in total, $55MN_r$ flops are required for generating soft outputs. 
In summary, the overall per-bit complexity for both detection and generating soft output is $\mathcal{O}(N_r)$.

\subsection{Soft Output MMSE Detector}
In the case of MMSE detector, it has been shown that the complexity for detection 
is $10N_r^3 + 5.5N_r^2 + 1.5N_r$ flops \cite{complexity_cal}. 
For generating of soft outputs, the likelihood of $\hat{s}_k$ conditioned on $s \in \mathbb{A}^{M}$ is given by \cite{Eq_AWGN},
\begin{equation}
\label{Prob_eq_MMSE}
\mathrm{P}\left( \hat{s}_k \mid s \right) =  \frac{1}{\sqrt{2\pi\epsilon^{2}_{k}}}~ \text{exp} \left( -\frac{1}{2\epsilon^{2}_{k}}\|\hat{s}_k - \mu_k s\|^{2}\right),
\end{equation}
where $\mu_k = \mathbf{W}_{\textrm{mmse},k}^{\mathsf{H}}\mathbf{H}_k$ which requires $4N_r - 1$ flops and  
$\epsilon^{2}_{k} = \frac{E_s}{N_t} ( \mu_k - \mu_{k}^{2} )$ which requires 3 flops.
Again, each transmitted symbol have $M$ level to be calculated and we have $N_t = N_r$ transmitted symbols.
To calculate soft output defined in (\ref{Prob_eq_MMSE}), 
$4MN_r^2 + 58M$ flops 
are required for generating soft outputs from all estimated symbols.

Table \ref{flop_need} summarises the computational complexity of both the proposed detector and the MMSE detector in terms of flops.
For each transmission in $200 \times 200$ MIMO system with BPSK modulation,
$80,540,416$ flops are required for the MMSE detector while
we need only $221,800$ flops for the proposed MF detector  (just 0.28 $\%$ to the MMSE detection)
Thus, by using the proposed detector, the computational complexity of large MIMO detection is strongly reduced.
\begin{table}[htb]
\setbox0\hbox{\verb/\documentclass/}%
\caption{The number of flops for MIMO detection.}
\label{flop_need}
\begin{center}
\begin{tabular}{| c | c |  c | }
 \hline
 Type $\diagdown$ Operation & Detection &  Soft Output \\
  \hline
Proposed detector & $5N_r^2 - N_r$ & $55MN_r$ \\
  \hline
MMSE detector & $10N_r^3 + 5.5N_r^2 + 1.5N_r$  & $4MN_r^2 + 58M$ \\
  \hline
\end{tabular}
\end{center}
\end{table}

\section{Conclusions}
In this paper, we develop the soft output detector 
based on MF technique for non-binary LDPC coded large MIMO systems.
An advantage of the proposed system is very low complexity of detection and soft output generation,
i.e., overall complexity is just 0.28$\%$ when comparing with that of the soft output MMSE detector.
By using 200 transmit/receive antennas and BPSK modulation, 
the performance within 4.2 dB from the capacity limit can be obtained at spectral efficiency below 66.67 bps/Hz.
Due to the low complexity detection, 
one can expect the improved performance by considering 
the joint iterative detection and decoding in turbo fashion.
The performance of the proposed coded system with higher modulation is also not shown in this paper.
Both will be further explored as the future work.

\section*{Acknowledgements}
This work is financially supported by the Telecommunications Research Industrial and Development Institute (TRIDI), 
with National Telecommunications Commission (NTC), Grant No.PHD/009/2552.

\bibliographystyle{IEEEtran}
\bibliography{IEEEabrv,my_references}

\end{document}